\documentclass[11pt,a4paper]{article}

\usepackage{jheppub}
\usepackage{amsthm}

\newcommand{\p}{\partial}

\newcommand{\N}{{\nabla}}
\newcommand{\Nb}{{\overline \nabla}}

\newcommand{\ad}{{\dot a}}
\newcommand{\bd}{{\dot b}}
\newcommand{\cd}{{\dot c}}
\newcommand{\lb}{{\bar\lambda}}
\newcommand{\tb}{{\bar\theta}}
\newcommand{\R}{{\bf R}}
\renewcommand{\a}{{\alpha}}

\renewcommand{\b}{{\beta}}

\renewcommand{\d}{{\delta}}
\newcommand{\e}{{\epsilon}}

\newcommand{\g}{{\gamma}}

\newcommand{\G}{{\Gamma}}
\renewcommand{\i}{{\iota}}
\renewcommand{\l}{{\lambda}}

\newcommand{\s}{{\sigma}}

\renewcommand{\t}{{\theta}}

\title{Geometrical framework for picture changing operators in the pure spinor formalism}
\author{Andrei Mikhailov}
\author{and Dennis Zavaleta}
\emailAdd{a.mikhaylov@unesp.br}
\emailAdd{dennis.zavaleta@unesp.br}

\affiliation{Instituto de Fisica Teorica, Universidade Estadual Paulista\\
R. Dr. Bento Teobaldo Ferraz 271, 
Bloco II -- Barra Funda\\
CEP:01140-070 -- Sao Paulo, Brasil\\
}

\abstract{
   It is well known in NSR string theory, that vertex operators
   can be constructed in various ``pictures''. Recently this was discussed
   in the context of pure spinor formalism.  NSR picture changing operators
   have an elegant super-geometrical interpretation. In this paper we
   provide a generalization of this super-geometrical construction, which is also
   applicable to the pure spinor formalism.
}

\begin{document}

\maketitle

\section{Introduction}\label{Introduction}

Pure spinor formalism is very promising for studying strings in $AdS$, since it naturally
includes the Ramond-Ramond fields \cite{Berkovits:2008ga,Berkovits:2008qc}.
However, the progress
has been slowed down by the lack of explicit formula
for vertex operators
(see \cite{Berkovits:2000yr,Mikhailov:2011si} for definitions and \cite{Bedoya:2010qz} for a simple example).
Recently, a promising new method was suggested in \cite{Berkovits:2019rwq}. The main idea is to
construct the vertex in nonzero picture.\footnote{
         Picture changing operators were previously discussed in the context of pure spinor formalism
         in \cite{Grassi:2004tv,Berkovits:2006vi}. More generally, the target space picture changing
         operators were essential in constructing the Chern-Simons actions in \cite{Cremonini:2019aao}.
         } This means allowing delta-functions of the pure
spinor ghosts. It was shown in \cite{Berkovits:2019rwq}, that the ansatz for 1/2-BPS vertex is simplified
in the -8 picture. Then,  the vertex operator in picture zero can be obtained by applying picture
raising operators, as explained in \cite{Martins:2019wwi}.
However, this picture-raising procedure is not the usual one \cite{Friedan:1985ge}, because the pure spinor variables
are constrained to live on a cone. In particular, it does not immediately fit into the geometrical
framework of \cite{Belopolsky:1997bg,Belopolsky:1997jz}.

In this paper we will develop a generalization of the approach of \cite{Belopolsky:1997bg,Belopolsky:1997jz}, which does cover the construction of \cite{Berkovits:2019rwq,Martins:2019wwi}.

The pure spinor target space can be considered a generalization of the odd tangent bundle $\Pi TX$
over the super-space-time $X$ (see Section \ref{sec:PiTM}).
The generalization consists of imposing some constraints on the
coordinates in the fiber. If $X$ has coordinates $x,\theta$, then $\Pi TX$ has coordinates
$x,\theta,dx,d\theta$. We consider a submanifold
$C\subset \Pi TX$ defined by some quadratic and linear equations --- see Section \ref{ConstrainedSurface}
and \cite{Movshev:2012gc}.
The idea is to construct the action of some ``odd loop group'' $\Pi TG$ on $C$ and then average over
its orbits, using the natural measure on $\Pi TG$.
The integration removes the delta-functions of the pure spinor ghosts, and the result is the picture
zero vertex operator.

One lesson from our study: it is useful to consider pure spinor vertex operators as,
roughly speaking, differential forms on space-time:
\begin{itemize}
\item vertex operators are (pseudo) differential forms
\end{itemize}
The pure spinor variable $\lambda$ is, in some sense,
$d\theta$ --- the differential of the fermionic coordinate of the target space, with the pure spinor
constraint imposed \cite{Movshev:2012gc}. The de Rham operator $d$ is identified with the BRST operator $Q$.
In this language picture changing operators are geometrical
operations on forms, basically averaging over an orbit of some group. We want this operation
to commute with the action of $Q = d$. To achieve that, we take the group to be of the form $\Pi TG$
for some Lie supergroup $G$, and construct its action in such a way that $d$ agrees with
the nilpotent vector field of $\Pi TG$, as explained in Section \ref{sec:UsualStory}.
The operation of averaging is usually non-local, an integral transform.
But it becomes local if all integrations are absorbed by delta-functions.
We do not want to allow delta-functions of the target space
coordinates, because we consider vertex operators  corresponding to smooth supergravity solutions.
Instead, we introduce delta-functions of $d\theta$, which means allowing ``pseudo-differential forms''.
It seems that our construction requires $G$ to be purely odd supergroup (in particular, abelian).
Basically, it is ${\bf R}^{0|8}$. If bosonic directions were present in $G$, then integration over them
would not be absorbed by delta-functions, leading to non-locality.

The main point of our approach is defining a structure of differential ${\bf R}^{0|8}$-module on
the space of vertex operators, Section \ref{SubmanifoldsInPiTM}. The construction is nontrivial,
and ``depends on some luck'', Section \ref{PiTRonC}. We hope that it would generalize to
the case of pure spinor superstring in AdS, but at this time we do not have such generalization.

It is also worth noting that this formalism has also been applied to the superstring in other dimensions (e.g. $D=3$ in \cite{Grassi:2004tv})
where there are no pure spinor constraints and so the delta functions on these variables does not present
any problem.

\emph{Plan of this paper.}
We first review the geometrical construction of \cite{Belopolsky:1997bg},\cite{Belopolsky:1997jz}
in Section \ref{GeometricalPictureChanging}, and discuss $d$-closed submanifolds in
Section \ref{SubmanifoldsInPiTM}.
Then we apply these concepts to pure spinor formalism in Section \ref{ConstrainedSurface}.
In Section \ref{PiTRonC} and Section \ref{AppAnticommute} we introduce a generalization of the
geometrical picture changing procedure of \cite{Belopolsky:1997bg,Belopolsky:1997jz} and
reproduce the result of \cite{Martins:2019wwi}. Finally, in Section \ref{OpenQuestions} we
discuss open questions.

\section{Geometrical interpretation of picture changing}\label{GeometricalPictureChanging}

\subsection{Reminder on odd tangent bundle $\Pi TX$}\label{sec:PiTM}

When studying a supermanifold $X$, it is often useful to consider, for any
``test'' supermanifold $S$, the space of maps:
\begin{equation}
   {\cal F}_X[S] = \mbox{Map}(S,X)
   \end{equation}
(see J. Bernstein's lectures in \cite{Deligne:1999qp}, which are also available online at

\texttt{https://www.math.ias.edu/QFT/fall/}).
This defines a contravariant functor $S\mapsto {\cal F}_X[S]$ from 
supermanifolds to sets.
This is the ``functor of points''; ${\cal F}_X[S]$ is called $S$-points of $X$.

If $X$ is a supermanifold, then functor $\Pi T$ in the category of supermanifolds is defined
as follows:
\begin{equation}
   \mbox{Map}(S, \Pi TX) = \mbox{Map}(S \times {\bf R}^{0|1}, X)
   \end{equation}
Functions on $\Pi TX$ are called ``pseudo-differential forms (PDFs) on $X$''. 

In particular, for a Lie supergroup $G$, the odd tangent space $\Pi TG$ is also
a Lie supergroup, which might  be considered an odd analogue of the loop group of $G$.
Let ${\bf g} = \mbox{Lie}(G)$ be the Lie algebra of $G$.
The Lie algebra of $\Pi TG$ is usually called ``cone of $\bf g$'' and denoted $C{\bf g}$:
\begin{equation}
   C{\bf g} = \mbox{Lie}\Pi TG
\end{equation}

\subsection{Picture raising operators}\label{sec:UsualStory}

Suppose that a Lie supergroup   $G$ acts on a supermanifold $X$.
Then $\Pi TG$ acts on $\Pi TX$. This means that pseudodifferential forms on $X$
form a linear representation of $\Pi TG$. Given a PDF $\omega \in \mbox{Fun}(\Pi TX)$
we consider:
\begin{equation}
   \Gamma\omega = \int_{g\in \Pi TG}  \omega\circ g
   \label{IntegralOverCg}
\end{equation}
We restrict ourselves to those $\omega$ for which this integral converges.
We observe that:
\begin{align}  
 &(\Gamma\omega)\circ g = 0 \quad\forall\quad g\in \Pi TG\label{Invariance} \\  
 &d\Gamma\omega = \Gamma d\omega\label{Differential} \end{align}
Eq. (\ref{Invariance}) implies that $\omega$ actually descends (as a PDF) on
the space of orbits of $G$ in $X$.
Eq. (\ref{Differential}) implies that closed $\omega$ gives closed $\Gamma\omega$.

This construction is usually applied to the case when $G$ is purely odd,
\textit{i.e.} take the simplest example $G = {\bf R}^{0|1}$ and $\Pi TG = {\bf R}^{1|1}$ with the target supermanifold $X$ parametrized by $(Z^M)$.
Suppose that $\omega$ contains enough delta-functions to absorb the integration along
odd variables.
In this case the convergence of the integral in  Eq. (\ref{IntegralOverCg}) is guaranteed,
and moreover $\omega\mapsto \Gamma\omega$ is actually a local operation. To be explicit,
suppose that the Lie algebra ${\bf g} = {\bf R}^{0|1}$ is generated by the odd vector fields $\nu$:
\begin{equation}
   \nu  = \nu^M(Z) \frac{\p }{\p  Z^M}
   \end{equation}
The action of the group $\R ^{0|1}\times  X\rightarrow  X$ is then given by:
\begin{equation}
   (\epsilon ,Z^M) \mapsto  Z^M + \e  \nu^M
   \end{equation}
The corresponding action of $\Pi TG$ on $\Pi TX$ is:
\begin{align}
  a\;:\;\Pi TG\times \Pi TX \;\rightarrow\; & \Pi TX
  \\  
  (\epsilon, d\epsilon , Z^M, dZ^M) \;\stackrel{a}{\mapsto}\; &
  (Z^M + \e \nu^M\,,\,dZ^M + d\epsilon \nu^M - \epsilon d\nu^M)
\end{align}
The integral of Eq. (\ref{IntegralOverCg}) is:
\begin{equation}
     (\Gamma_\nu  \omega)(Z,dZ) = \int D(d\e ) D(\e )
     \ \omega(Z^M +\e  \nu^M , dZ^M + d\e  \nu^M  - \e  d\nu^M )
\end{equation}
For any supermanifold $M$, the odd tangent space  $\Pi T M$ has a canonical vector field $d$.
If we thing of   $\Pi T M$ as the space of maps ${\bf R}^{0|1}\rightarrow M$, then this $d$ is the 
infinitesimal shift in ${\bf R}^{0|1}$. Eq. (\ref{Differential}) can be derived by observing that for
any $\omega$, a function on $\Pi TX$:
\begin{equation}\label{DifferentialAction}
(d_{\Pi TG} + d_{\Pi TX}) (\omega\circ a) = (d_{\Pi TX}\omega)\circ a
\end{equation}

After integrating on the odd parameter $\e$, we can arrive at the ``usual'' (in string theory literature)
expression for the PCO:
\begin{equation}
   (\Gamma_{\nu}  \omega) = [d, \Theta (\i _\nu )]
   \end{equation}
where $\iota_\nu = \nu^M \frac{\p}{\p dZ^M}$.

This can be generalized to $G = {\bf R}^{0|n}$ and $\Pi TG = {\bf R}^{n|n}$ where ${\bf g} = {\bf R}^{0|n}$ is generated by $n$ odd vector fields $\nu_a$ with $a=1,\ldots,n$. Then the integral becomes
  \begin{equation}\label{integral}
  (\Gamma_\nu  \omega)(Z,dZ) = \int \prod_a D(d\e^a ) \prod_a D(\e^a )\ \omega( g^{\e^a, d\e^a}(Z,dZ) )
  \end{equation}
where $g^{\e^a, d\e^a}$ is the flow corresponding to the infinitesimal transformation
  \begin{equation}
  (\epsilon^a, d\epsilon^a , Z^M, dZ^M) \mapsto (Z^M + \e^a  \nu_a ^M\,,\,dZ^M + d\epsilon^a \nu_a^M - \epsilon^a d\nu_a^M)
  \end{equation}

\subsection{Some properties of $\Pi T$}\label{sec:DefPiT}

Some computations are simplified (see Section \ref{PiTRonC}) by considering the
iterated application of $\Pi T$. Consider
\begin{equation}\label{FlatMap}
   \mu\;:\;\Pi T\Pi T X \rightarrow \Pi TX
   \end{equation}
induced by the diagonal map
\begin{equation}
   \Delta \;:\; {\bf R}^{0|1}\longrightarrow {\bf R}^{0|1}\times {\bf R}^{0|1}
   \end{equation}
inducing
  \begin{equation}
  \begin{aligned}
  \mu[S]\;:\;\mbox{Map}(S \times {\bf R}^{0|1}\times {\bf R}^{0|1}, X)
                \; & \longrightarrow\;\mbox{Map}(S \times {\bf R}^{0|1}, X) \\
  \phi\quad & \longmapsto \mu[S](\phi) \;=\; \phi\circ ({\rm id}\times \Delta)
  \end{aligned}
  \end{equation}
There is a canonical nilpotent odd vector field $d\in\mbox{Vec}(\Pi TX)$
generating shifts along ${\bf R}^{0|1}$.

For any odd vector field $Q\in(\mbox{Vec}(X))_{\bar{1}}$ we define the odd flux map
  \begin{align} g_Q[S]\;:\;
  &\mbox{Map}(S,X)\longrightarrow \mbox{Map}(S\times {\bf R}^{0|1},X)
  \end{align}
and an even vector field $\iota_Q\in(\mbox{Vec}(\Pi TX))_{\bar{0}}$ with the flux:
  \begin{equation}
  \begin{aligned}
  \exp(\iota_Q)[S]\;:\; &\mbox{Map}(S\times {\bf R}^{0|1},X) \longrightarrow \mbox{Map}(S\times {\bf R}^{0|1},X) \\
  \exp(\iota_Q)[S]\;=\; &\mu[S]\circ g_Q[S\times {\bf R}^{0|1}]
  \end{aligned}
  \end{equation}
And $l_Q = [\iota_Q,d]$.
                            
If we denote the coordinates of $\Pi T \Pi T X$ as $x,dx, Dx, Ddx$, the projection $\mu$ is:
\begin{equation}
   (f\circ \mu)(x,dx, Dx, Ddx) = f(x, dx + Dx)
   \end{equation}

\section{Submanifolds in $\Pi TX$}\label{SubmanifoldsInPiTM}

Consider a submanifold $C\subset \Pi TX$ which is closed under $d$.
(This means that the ideal generated by PDFs vanishing on $C\subset \Pi TX$ is
closed under $d$.) We explained in Section \ref{sec:UsualStory} that
the action of the group $\Pi TG$ on $\Pi TX$ can be obtained from the action
of $G$ on $X$. This, however, would not work for us here, because the resulting action
of $\Pi TG$ would not preserve $C\subset \Pi TX$.
But in fact, the validity of Eqs. (\ref{Invariance}) and (\ref{Differential}) does not
depend on \emph{how} we constructed the action of $\Pi TG$ on $\Pi TX$. For any action
of $\Pi TG$ on $\Pi TX$, commuting with the action of $d$ in the sense of Eq. (\ref{DifferentialAction}), the transformation
$\omega \mapsto \Gamma\omega$ defined
by Eq. (\ref{IntegralOverCg}) will satisfy Eqs. (\ref{Invariance}) and (\ref{Differential}).
This is equivalent to defining the structure
of a differential ${\bf g}$-module\footnote{A differential ${\bf g}$-module is a representation $V$
  of $C{\bf g}$, equipped with a differential $d_V$, compatible with the differential $d$ of $C{\bf g}$}
on the space of functions on $C$.
We will define the action of $C{\bf g}$ on $\Pi TX$ compatible with $d$
such that the vector fields representing $C{\bf g}$ will be tangent to $C\subset \Pi TX$.
This defines the structure of a differential ${\bf g}$-module on $\mbox{Fun}(C)$.
Then, we will define $\Gamma$ by Eq. (\ref{IntegralOverCg}).

\section{Pure spinor target space as a subspace in $\Pi TX$}\label{ConstrainedSurface}

\subsection{Supersymmetry generators and invariant derivatives}\label{sec:SUSY}

It is always possible
to find coordinates $x,\theta$ such that the supersymmetry generators have the form:
  \begin{equation}
   q^L_\a  = \frac{\p }{\p \t _L^\a }  - (\t _L\g ^m)_\a  \frac{\p }{\p  x^m} \  ,\qquad  q^R_\a  = \frac{\p }{\p \t _R^\a }  - (\t _R\g ^m)_\a  \frac{\p }{\p  x^m}
   \end{equation}
To construct supersymmetry-invariant objects, it is useful to know the vector
fields commuting with $q^{L|R}$. Besides translations $\frac{\partial}{\partial x^m}$,
there are  fermionic vector fields commuting  with $q^{L|R}$. They are:
  \begin{equation}
   \N ^L_\a  = \frac{\p }{\p \t _L^\a }  + (\t _L\g ^m)_\a  \frac{\p }{\p  x^m} \  ,\qquad  \N ^R_\a  = \frac{\p }{\p \t _R^\a }  + (\t _R\g ^m)_\a  \frac{\p }{\p  x^m}
   \label{DefNabla}\end{equation}
The only non-zero commutators are
  \begin{align}
  [\N ^L_\a , \N ^L_\b ] & = [\N ^R_\a ,\N ^R_\b ] = 2\g ^m_{\a \b } \frac{\p }{\p  x^m} \\
  [q^L_\a , q^L_\b ] & = [q^R_\a , q^R_\b ] = -2\g ^m_{\a \b } \frac{\p }{\p  x^m}
  \end{align}

\subsection{Description of $C\subset \Pi TX$}\label{sec:IdealInPiTX}

Consider the space $X$ parametrized by the coordinates $(x^m, \t _L^\a , \t _R^\a )$.
Then $\Pi TX$ is pa\-ra\-me\-tri\-zed by the coordinates  $(x^m, \t _L^\a , \t _R^\a, dx^m, d\theta_L^\a, d\theta_R^\a )$ where ``$dx$'', ``$d\theta$'' are considered one letter.
Then, to describe the pure spinor string in a flat background, we choose as a target the subspace of $C\subset \Pi  T X$ defined by the following conditions
\begin{align}  
 &d\t _L \g ^m d\t _L = d\t _R \g ^m d\t _R = 0\label{cond1} \\  
 &d x^m - d\t _L \g ^m\t _L - d\t _R \g ^m\t _R = 0\label{cond2} \end{align}
The constraint defined by Eq. (\ref{cond1}) is the pure spinor constraint,
it is essentially postulated. The constraint of Eq. (\ref{cond2}) is characterized by:
\begin{align}  
 &\iota_{\nabla^{L|R}}(d x^m - d\t _L \g ^m\t _L - d\t _R \g ^m\t _R)\;=\;0\label{IotaNablaAndDX} \end{align}
where $\nabla^{L|R}$ are defined by Eq. (\ref{DefNabla}). 

The main properties of the ideal generated by Eqs. (\ref{cond1}) and (\ref{cond2}) are:
\begin{enumerate}
\item It is $q^L$- and $q^R$-invariant, {\it i.e.} ${\cal L}_{q^{L|R}}$ annihilate\footnote{
To see the vanishing of ${\cal L}_{q^{L|R}}(d x^m - d\t _L \g ^m\t _L - d\t _R \g ^m\t _R) =0$,
we first observe that this expression does not contain $dx$. But then, it cannot
contain neither $d\t _L$ nor $d\t _R$ because of Eq. (\ref{IotaNablaAndDX}).}
  $d\t _L \g ^m d\t _L$, $d\t _R \g ^m d\t _R$, and $d x^m - d\t _L \g ^m\t _L - d\t _R \g ^m\t _R$
\item It is $d$-closed.
\end{enumerate}

Usually one denotes:
\begin{equation}
   \lambda_L^{\alpha} = d\theta^{\alpha}_L\,,\quad \lambda_R^{\alpha} = d\theta^{\alpha}_R
   \end{equation}
The BRST operator $Q$ is just $d$:
  \begin{equation}
   \begin{aligned}     
  Q(x^m) & = (\l _L\g ^m\t _L) + (\l _R\g ^m\t _R) \\
  Q(\t ^\a _L) & = \l ^\a _L \\
  Q(\t ^\a _R) & = \l ^\a _R \\
  Q(\l ^\a _L) & = 0 \\
  Q(\l ^\a _R) & = 0
  \end{aligned}
   \end{equation}

Eq. (\ref{IotaNablaAndDX}) is promising for constructing the action of
$\Pi TG$ on $\Pi TX$ as described in Section \ref{SubmanifoldsInPiTM}.
We also observe:
\begin{equation}
   {\cal L}_{\nabla^{L|R}}(d\theta_{L|R}\gamma^m d\theta_{L|R}) = 0
                   \end{equation}
However, let us keep in mind that:
  \begin{align}
  &\iota_{\nabla^{L|R}}(d\theta_{L|R}\gamma^m d\theta_{L|R}) \neq 0 \\
  &{\cal L}_{\nabla^{L|R}}(d x^m - d\t _L \g ^m\t _L - d\t _R \g ^m\t _R)\neq 0
  \end{align}

\subsection{Solving the pure spinor constraint}\label{sec:SolvingPSConstraint}

The ten-dimensional gamma matrices can be decomposed to give the eight-dimensional ones and also chiral projectors
\begin{equation}
   \begin{aligned}     
   {}[\g ^i_{\a \b }] & =
   \left (
          \begin{array}{cc}     
          0 & \s ^i_{a\bd } \\
          \s ^i_{\ad  b} & 0
          \end{array} 
          \right )\  ,\quad  [\g ^{i\a \b }] =
   \left (
          \begin{array}{cc}     
          0 & \s ^{ia\bd } \\
          \s ^{i\ad  b} & 0
          \end{array} 
          \right ) \\ 
  [\g ^+_{\a \b }] & =
  \left (
         \begin{array}{cc}     
         0 & 0 \\
         0 & -\d _{\ad \bd }
         \end{array} 
         \right )\  ,\quad  [\g ^{+\a \b }] =
  \left (
         \begin{array}{cc}     
         \d ^{ab} & 0 \\
         0 & 0
         \end{array} 
         \right ) \\ 
  [\g ^-_{\a \b }] & =
  \left (
         \begin{array}{cc}     
         -\d _{ab} & 0 \\
         0 & 0
         \end{array} 
         \right )\  ,\quad  [\g ^{-\a \b }] =
  \left (
         \begin{array}{cc}     
         0 & 0 \\
         0 & \d ^{\ad \bd }
         \end{array} 
         \right )
\end{aligned}
   \end{equation}
where $i,j,a,b,\ad ,\bd  = 1,\ldots , 8$. Then, we will use the Kronecker deltas $\d _{ab},\d _{\ad \bd },\d ^{ab},\d ^{\ad \bd }$ to raise and lower spinor indices. The Pauli matrices are such that $\s ^i_{a\bd } = \s ^i_{\bd  a}$ and $(\s ^j)^{\ad  b} = \d ^{\ad  \bd }\d ^{b a}\s ^j_{a\bd }$. These matrices satisfy
  \begin{align}  
  &(\s ^i)_{a\bd } (\s ^j)^{\bd  c} + (\s ^j)_{a\bd }(\s ^i)^{\bd  c} = 2 \d _a^c \d ^{ij} \\  
  &(\s ^i)_{a\bd } (\s _i)_{c \dot  d} + (\s ^i)_{a\dot  d} (\s _i)_{\bd  c} = 2 \d _{ac}\d _{\dot  d \bd }\label{fierz} \\  
  &(\s ^{ij})^a{}_b(\s _{ij})^c{}_d = 8 \d ^{ac}\d _{bd} - 8 \d ^a_d \d ^c_b \\  
  &(\s ^{ij})^a{}_b(\s _{ij})^\cd _{\dot  d} = 4 (\s ^i)^{a\cd }(\s _i)_{b\dot  d} - 4 \d ^a_b\d ^\cd _{\dot  d}\label{pauli-identity}
  \end{align}
In $SO(8)$ components, the pure spinor constraints for $\l ^\a  = (\l ^a,\lb ^\ad )$ are:
\begin{equation}
   \l ^a\d _{ab} \l ^b = 0\  ,\quad  \lb ^\ad \d _{\ad \bd } \lb ^\bd  = 0\  ,\quad  \l ^a\s ^i_{a\ad }\lb ^\ad  = 0
   \label{ps-cond}\end{equation}
Both $\lambda_L$ and $\lambda_R$ satisfy these constraints.

They can be solved as follows. Let us define:
\begin{equation}
   \l _\pm ^a := \frac{1}{2} (\l _L^a \pm  i \l _R^a)\  ,\quad  \lb _\pm ^\ad  := \frac{1}{2} (\lb _L^\ad  \pm  i \lb _R^\ad )
   \end{equation}
Eqs. (\ref{cond1}) imply:
  \begin{align}  
  &\l_+^2 + \l_-^2 = \l_+\l_- = 0 \\  
  &\lb_+^2 + \lb _-^2 = \lb _+\lb _- = 0 \\  
  &\l_+\s ^i\lb _+ + \l _-\s ^i\lb _- = \l _+\s ^i\lb _- + \l _-\s ^i\lb _+ = 0 
  \end{align}
We can solve these equations for $\l _-^a$ in terms of the rest of variables:
\begin{equation}
   \l _-^a = - \frac{(\lb _+\s _{ij}\lb _-)}{4(\lb _+\lb _+)}  (\s ^{ij})^a{}_b \l _+^b = \frac{(\lb _L \s ^{ij}\lb _R)}{4(\lb _L \lb _R)}  (\s _{ij})^a{}_b\l _+^b
   \label{LambdaMinusAsAFunctionOf}
\end{equation}
where $\l _+^a$ is unconstrained and $(\lb ^\ad _L$,$\lb ^\ad _R)$ are still subject to the conditions $(\lb _L)^2 = (\lb _R)^2 = 0$.
Let us introduce the  $8\times 8$ matrix $M$:
\begin{equation}\label{DefM}
  M^a{}_b = {\partial\lambda^a_- \over\partial\lambda^b_+} =
  \frac{(\overline\lambda_L\sigma^{ij}\overline\lambda_R)}{4(\overline\lambda_L\overline\lambda_R)}(\sigma_{ij})^a{}_b
\end{equation}
This matrix inherits the antisymmetry properties of the Lorentz generators $(\sigma_{ij})^a{}_b$ when their indices are raised and lowered with Kronecker deltas\\
\begin{equation}
M_{ab} = -M_{ba}\quad\text{equivalently}\quad M^a{}_b = -M_b{}^a\\
\end{equation}
and it squares to identity:
\begin{equation}\label{SquareOfM}
M^a{}_b M^b{}_c = \delta^a_c
\end{equation}
(Eq. (\ref{SquareOfM}) follows from considering separately the identity    $(\lambda_+)^2 + (\lambda_-)^2 = 0$.)

In the light-cone coordinates the conditions (\ref{cond2}) are written as
  \begin{align}  
  &dx^+ - 2 \left ( \l _+ \t _- + \l _- \t _+ \right ) = 0\label{constraint1} \\  
  &dx^- - 2 \left ( \lb _+ \tb _- + \lb _-\tb _+ \right ) = 0 \label{constraint2}\\  
    &dx^i - 2 \left ( \l _+\s ^i\tb _- + \l _-\s ^i\tb _+ + \lb _+\s ^i\t _- + \lb _-\s ^i\t _+ \right ) = 0
    \label{constraint3}
  \end{align}
Consider the subspace of functions which only depend on $x^+$ and do not on $x^-$ nor $x^i$.
On this subspace:
  \begin{align}  
  &\N ^\pm _a = \frac{\p }{\p  \t _\pm ^a}  + 2 \t _{\mp  a} \frac{\p }{\p  x^+} \  ,\quad  \Nb ^\pm _\ad  = \frac{\p }{\p  \tb _\pm ^\ad } \\  
  &q^\pm _a = \frac{\p }{\p  \t _\pm ^a}  - 2 \t _{\mp  a} \frac{\p }{\p  x^+} \  ,\quad  \bar  q^\pm _\ad  = \frac{\p }{\p  \tb _\pm ^\ad }
  \end{align}
while the only non-zero commutators will be $[\N ^+_a, \N ^-_b] = - [q^+_a, q^-_b] = 4\d_{ab} \p _+$.

\section{An action of $C {\bf R}^{0|8}$ on $C$}\label{PiTRonC}

Let us consider the coordinates on the fiber of $C$:
$(\lambda^a_+,\bar{\lambda}^a_+,\bar{\lambda}^a_-)$.
Let us introduce the following vector fields on $C$:
\begin{equation}
   i_a =  \frac{\partial}{\partial\lambda^a_+}
   \label{VectorFieldWithSolvedConstraint}\end{equation}
These $i_a$ are vertical vector fields (tangent to the fiber).
They commute: $[i_a,i_b] = 0$.
         Equivalently, we can start with unconstrained $\lambda$, and define:
         \begin{equation}\label{IaExpanded}
           i_a = {\frac{\partial}{\partial\lambda_+^a}} +
           \left.{\frac{\partial\lambda_-^b}{\partial\lambda_+^a}}\right|_{\bar{\lambda}=const} {\frac{\partial}{\partial\lambda_-^b}}
           +
           \left.{\partial dx^m\over\partial \lambda_+^a}\right|_{\bar{\lambda}=const} {\partial\over\partial dx^m}
           +
           \left.{\partial dx^+\over\partial \lambda_+^a}\right|_{\bar{\lambda}=const} {\partial\over\partial dx^+}
         \end{equation}
         where the derivative $\left.{\frac{\partial\lambda_-^b}{\partial\lambda_+^a}}\right|_{\bar{\lambda}=const}$ is of the RHS of Eq. (\ref{LambdaMinusAsAFunctionOf}), and ${\partial dx\over\partial \lambda_+^a}$ includes
         the explicit dependence of $dx$ on $\lambda_+$ as well as dependence through $\lambda_-(\lambda_+)$
         (see Eqs. (\ref{constraint1}), (\ref{constraint2}), (\ref{constraint3})).
         This vector field is tangent to the cone.
         This is the same as to consider the vector field of Eq. (\ref{VectorFieldWithSolvedConstraint}) on $C$
         in coordinates $(\lambda^a_+,\bar{\lambda}^a_+,\bar{\lambda}^a_-)$.

Next, we define:
\begin{equation}
   l_a = [i_a,d] = \Pi T(\N ^+_a) + \left.{\frac{\partial\lambda_-^b}{\partial\lambda_+^a}}\right|_{\bar{\lambda}=const}  \Pi T(\N ^-_b)
   \label{ExplicitLa}\end{equation}
In deriving Eq. (\ref{ExplicitLa}), the following observation is useful.
If $\upsilon$ is an even vertical vector field on $\Pi TX$, then:
\begin{equation}
   \mu_* \left(\iota_{\Pi T\Pi TX}([\upsilon, d_{\Pi TX}])\right)
   = \upsilon
\end{equation}
where $\mu$ is from Eq. (\ref{FlatMap}).
Indeed, the RHS of Eq. (\ref{ExplicitLa}) is fixed by $\mu_* \left(\iota_{\Pi T\Pi TX}l_a\right)$ being
as in Eq. (\ref{IaExpanded}) (this follows from the definition of $\nabla$, Eq. (\ref{IotaNablaAndDX})).

We observe that $\lambda_-^b$ is a linear function of $\lambda_+^a$, thus
${\frac{\partial^2\lambda_-^a}{\partial\lambda_+^b\partial\lambda_+^c}} = 0$. Also
${\partial \lambda_-^b\over\partial \lambda_+^a} + {\partial \lambda_-^a\over\partial \lambda_+^b} = 0$.
Therefore:
\begin{equation}[l_a, i_b] = 0\end{equation}
and this implies
\begin{equation}
   [l_a,l_b] = 0
   \end{equation}
This means that $d,i_a,l_b$ define an action of the differential Lie superalgebra
$C{\bf R}^{0|8}$ on $C$.

\subsection{Type IIB supergravity vertex operator at picture $(-8)$}\label{sec:VO8}

In coordinates $y^+ = x^+ - 2\t _+^a \t _{-a}$:
  \begin{align}
  \N _{+a} &=\frac{\p }{\p \t _-^a}  + 4 \t _{+a}\frac{\p }{\p  y^+} \  ,\quad  \N _{-a} = \frac{\p }{\p \t _+^a} \\
  q_{+a} &=\frac{\p }{\p \t _-^a} \  ,\quad  q_{-a} = \frac{\p }{\p \t _+^a}  - 4 \t _{-a} \frac{\p }{\p  y^+}
  \end{align}
We have lowered their $+,-$ superscripts such that $\nabla_{\pm a} := \nabla_a^\mp$ and $q_{\pm a} := q_a^\mp$. These coordinates simplify the $\t$-expansion of the di\-la\-ton su\-per\-field $\Phi (y^+,\t _\pm )$ since it satisfies $\N _- \Phi  = 0$. We have
  \begin{align}  
  &\Phi (y^+,\t _-) \;=\;\nonumber{} \\ \;=\;
  &e^{ik_+ y^+} \left( C + C^{a_1} \t _{-a_1} + \frac{1}{2!}  C^{a_1 a_2} \t _{-a_1} \t _{-a_2} + \ldots  + \frac{1}{8!}  C^{a_1\ldots  a_8} \t _{-a_1}\ldots  \t _{-a_8}\right)
  \end{align}
where the constants $C^{a_1\ldots  a_k}$ are bosonic (fermionic) when $k$ is even (odd). 

When working with the vertex operator corresponding to type IIB supergravity is useful to know the following identity
\begin{align}  
 &\prod _{b=1}^8 \N _{+b} \left ( e^{ik_+ y^+} \t _-^{a_1}\ldots  \t _-^{a_k} C_{a_1\ldots  a_k} \right ) \;=\;\nonumber{} \\ \;=\;
 &(4\p _+)^{8-k} C^{a_1\ldots  a_k} q_{-a_1}\ldots  q_{-a_k}\left ( e^{ik_+ y^+} \prod_{b=1}^8 \t _+^b \right )\end{align}
which in turn implies that
  \begin{align}  
  &\frac{1}{(4\p _+)^8} \prod _{b=1}^8 \N _+^b \Phi (y^+,\t _-) \;=\; \nonumber \\
  \;=\; &\left[ \sum _{k=0}^8 \frac{(4\p _+)}{k!} ^{-k} C^{a_1\ldots  a_k} q_{-a_1}\ldots  q_{-a_k}\right ]  \left ( e^{ik_+ y^+} \prod_{b=1}^8\t ^b_+ \right ) \label{full-multiplet}
  \end{align}
We start by considering the vertex operator in a $(-8)$-picture that corresponds to the type IIB supergravity scalar state
  \begin{equation}
   V_{-8}^{scalar} = (\lb _L \lb _R)e^{ik_+ y^+} \prod _{a=1}^8 \t ^a_+ \d (\l ^a_+)
   \end{equation}
and by applying the $q_{-a}$ supersymmetry generators, we can generate the full type IIB supergravity multiplet using (\ref{full-multiplet})
  \begin{equation}
   V_{-8} = (\lb _L \lb _R) \frac{1}{(4\p _+)^8} \left (\prod _{a=1}^8 \d (\l _+^a) \N _+^a\right ) \Phi (y^+,\t _-)
   \end{equation}

\subsection{Type IIB supergravity vertex operator at picture $(0)$}\label{sec:VO0}

We will now use the vector fields $\rho(l_a)$, $\rho(i_a)$ constructed in Section \ref{PiTRonC} to
transform the supergravity vertex from picture -8 to picture 0. As we explained in
Section \ref{sec:UsualStory} and Section \ref{SubmanifoldsInPiTM}, this amounts to computing the integral:
\begin{equation}
   V_0 := \Gamma  V_{-8}
   =\int  \prod _{a=1}^8 D(d\e ^a) \frac{\p }{\p \e ^a}  \exp \left ( \e ^a\rho (l_a) + d\e ^a\rho (i_a) \right ) V_{-8}
   \end{equation}
The explicit expressions for these vector fields are:
  \begin{align}
  \rho (l_a) &=\Pi  T(\N ^+_a) + \frac{\p \l _-^b}{\p \l _+^a}  \Pi  T(\N ^-_b) = \left ( \N ^+_a - 2 \l _{-a} \frac{\p }{\p  dx^+}  \right ) + \frac{\p \l _-^b}{\p \l _+^a}  \left ( \N ^-_b - 2 \l _{+b} \frac{\p}{\p dx^+} \right ) \\
  \rho (i_a) &=\left ( \frac{\p }{\p \lambda _+^a}  + 2\theta _-^a \frac{\p }{\p  dx^+}  \right ) + \frac{\p \l _-^b}{\p \l _+^a}  \left ( \frac{\p }{\p \l _-^b}  + 2\theta _+^b{\frac{\partial}{\partial dx^+}} \right ) \\
  \rho (d) &= d
  \end{align}
After some computation we can verify that these vector fields satisfy
  \begin{equation}
   [\rho (l_a),\rho (l_b)] = [\rho (i_a), \rho (i_b)] = [\rho (l_a), \rho (i_b)] = 0
   \end{equation}
Now we can integrate along the orbits of $\Pi T {\bf R}^{0|8}$, as in Eq. (\ref{integral}):
  \begin{align}
  V_0 := \G  V_{-8}
  &=\int  \prod _{a=1}^8 D(d\e ^a) \frac{\p }{\p \e ^a}  \exp \left ( \e ^a\rho (l_a) + d\e ^a\rho (i_a) \right ) V_{-8}\;=\;\nonumber{} \\  
  &=\int  \prod _{a=1}^8 D(d\e ^a) \rho (l_1)\ldots  \rho (l_8) \exp \left ( d\e ^a\rho (i_a)\right ) V_{-8}\;=\;\nonumber{} \\  
  &=\frac{1}{(4\p _+)^8} (\lb _L \lb _R) \rho (l_1)\ldots  \rho (l_8) \left ( \prod _{a=1}^8 \N _+^a \right ) \Phi (x^+ - 2\t ^a_+ \t _{-a},\t _-) 
  \end{align}
where in the last line we have used that the only $\l _\pm$-dependence of $V_{-8}$ is through $\d (\l _+^a)$ which means that the integral on $d\e$ eliminates all deltas at once. To proceed with the computation first notice that there is no dependence on coordinate $dx^+$, so we can drop the $\frac{\p }{\p  dx^+}$-part in $\rho (l_a)$. We also change to coordinates $y^+ = x^+ - 2\t _+^a \t _{-a}$ to obtain
\begin{align} \G  V_{-8}\;=\;
 &\frac{1}{(4\p _+)^8} (\lb _L\lb _R) \prod _{a=1}^8 \left ( \N ^+_a + \frac{\p \l _-^b}{\p \l _+^a}  \N ^-_b \right ) \left (\prod _{b=1}^8 \N _+^b\right ) \Phi (y^+,\t _-)\nonumber{} \\ \;=\;
 &\frac{1}{(4\p_+)^8}(\lb_L\lb_R) \e^{a_1\ldots a_8} \left[ \frac{1}{8!} \N^+_{a_1}\ldots \N^+_{a_8} + \frac{(4\p_+)}{6!2!} \N^+_{a_1}\ldots \N^+_{a_6} \frac{\p \l_{-a_8}}{\p\l^{a_7}_+} \right .\nonumber{} \\  
 &+ \frac{(4\p_+)^2}{4!(2!)^3} \N^+_{a_1} \N^+_{a_2} \N^+_{a_3} \N^+_{a_4} \frac{\p \l_{-a_6}}{\p\l^{a_5}_+} \frac{\p \l_{-a_8}}{\p\l^{a_7}_+}  + \frac{(4\p_+)^3}{3!(2!)^4} \N^+_{a_1} \N^+_{a_2} \frac{\p \l_{-a_4}}{\p\l^{a_3}_+} \frac{\p \l_{-a_6}}{\p\l^{a_5}_+} \frac{\p \l_{-a_8}}{\p\l^{a_7}_+}\nonumber{} \\  
 &\left. + \frac{(4\p_+)^4}{4!(2!)^4} \frac{\p \l_{-a_2}}{\p\l^{a_1}_+} \frac{\p \l_{-a_4}}{\p\l^{a_3}_+} \frac{\p \l_{-a_6}}{\p\l^{a_5}_+} \frac{\p \l_{-a_8}}{\p\l^{a_7}_+} \right] \left(\prod_{b=1}^8 \N_+^b\right) \Phi(y^+,\t_-)\label{GammaVMinus8} \end{align}
We want to further transform this expression.
First, we substitute Eq. (\ref{LambdaMinusAsAFunctionOf}) for
$\frac{\p \l _-}{\p \l _+}$.
Then, we anticommute all  $\N ^+$'s all the way to right and make use of:
\begin{equation}
   \N ^+\Phi =0
   \end{equation}
(Also, remember that $\nabla_a^- = \nabla_{+a}$ and $\nabla_a^+ = \nabla_{-a}$).
The computation uses some identities for the commutators of the SUSY-invariant derivatives,
namely Eqs. (\ref{expr-1}), (\ref{expr-2}) and (\ref{expr-3}) in Appendix Section \ref{AppAnticommute}.
The result is the following expression for the type IIB supergravity vertex operator $V_0 = \G  V_{-8}$:
  \begin{align} V_0
 &= (\lb _L\lb _R) \left[ \Phi  + \frac{1}{(32\p _+)}  \frac{(\lb _L \s ^{ij}\lb _R)}{(\lb _L\lb _R)} (\N _+ \s _{ij}\N _+) \Phi  \right .\nonumber{} \\  
 &\qquad \qquad + \frac{1}{2!(32\p_+)^2} \prod_{n=1}^2\frac{(\lb_L \s^{i_n j_n}\lb_R)}{(\lb_L\lb_R)} (\N_+ \s_{i_n j_n}\N_+) \Phi\nonumber{} \\  
 &\qquad \qquad + \frac{1}{3!(32\p_+)^3} \prod_{n=1}^3 \frac{(\lb_L \s^{i_nj_n}\lb_R)}{(\lb_L\lb_R)}(\N_+ \s_{i_n j_n}\N_+) \Phi\nonumber{} \\  
 &\qquad \qquad \left. + \frac{1}{4!(32\p_+)^4} \prod_{n=1}^4 \frac{(\lb_L \s^{i_nj_n}\lb_R)}{(\lb_L\lb_R)}(\N_+ \s_{i_n j_n}\N_+) \Phi \right]\label{V0} \end{align}
  This formula  reproduces the result of \cite{Martins:2019wwi} where the computation was performed using the PCOs in their standard form $\Gamma _a = [Q_{BRST}, \Theta (\omega ^+_a)]$.
  Our computation is, in some sense, more streamlined. In particular, we do not have any poles in $\lambda _+^a$. (The computation of \cite{Martins:2019wwi} has such poles at the middle steps; they appear
  each time a $\Theta (\omega ^+_a)$ hit a $\delta (\lambda ^a_+)$.)

  Still, our computation does suffer from the poles with denominators
  $(\bar \lambda _L \bar \lambda _R)$. The only terms which could potentially have such denominators
  are the third, fourth and fifth. Notice, however, that the fifth term is $(\lb_L \lb_R)$ times
  an expression which contains $\lb_L$ and $\lb_R$ only through
  the Pfaffian of the matrix $M$ defined in Eq. (\ref{DefM}). Since $M^2={\bf 1}$, the Pfaffian
  is $\pm 1$. Therefore, this last term only depends on $\lb_L$ and $\lb_R$ through the
  factor $(\lb_L \lb_R)$. It is equal to:
  \begin{equation}
    \pm (\lb_L \lb_R)\frac{1}{(16\p_+)^4}  \nabla_+^1\nabla_+^2\cdots\nabla_+^8 \Phi
  \end{equation}
  The fourth term (the one containing six $\nabla_+$) is also nonsingular. Indeed, it follows from
  $\mbox{det}(M)=1$ that its dependence on $\lb_L$ and $\lb_R$ is linear in $(\lb_L \lb_R)M$.
  The only term which could potentially have a pole is the third (middle) term, which contains four $\nabla_+$.
  It could have a first order pole in $(\lb_L \lb_R)$. The cancellation of this pole is not obvious;
  it was proven in \cite{Martins:2019wwi} using certain identities for the $(\sigma ^j)_{a\dot  a}$ matrices.
  
  Thus the zero-picture vertex operator is actually
  of the form $V_0 = \bar \lambda _L^{\dot  a} \bar \lambda _R^{\dot  b} A_{\dot  a \dot  b}(x^+,\theta _\pm )$.

\section{Open Questions}\label{OpenQuestions}

\begin{enumerate}
        \item 
        We constructed \emph{some} action of $C{\bf R}^{0|8}$ on the pure spinor target space.
        How flexible is this construction? Is it in some sense natural?
        Given $V_{-8}$, is there a \emph{canonical} way to construct $V_0$?
        (This question is very important for the computation of amplitudes.
              If there is no \emph{canonical} map $V_{-8} \rightarrow V_0$, then
              using the -8 picture in computation of amplitudes is not apriori justified.)
        \item 
        It is not immediately clear how to apply our method in AdS.
        A naive analogue of our vector fields $i_a$ and $l_a$ from Section \ref{PiTRonC}
        is not well-defined in AdS (it would not be gauge invariant).
        \item 
        The use of delta-functions on the cone is potentially dangerous.
        We would want to understand, when generalized functions of the form of
        products of $\delta(\lambda)$ are well-defined.
        \end{enumerate}

\appendix

\section{Anticommuting $\N_-$ and $\N_+$}\label{AppAnticommute}
Here we will prove some formulas of SUSY-invariant derivative $\nabla$ which
are needed to derive Eq. (\ref{V0}) from Eq. (\ref{GammaVMinus8}).

There are two relations that are useful when anticommuting the $\N _-$ and $\N _+$ operators. The first one which can be proven by induction is
\begin{align} \N _{-a_k}\ldots  \N _{-a_1} \N _+^{a_1}\ldots  \N _+^{a_k}
 &= (\N _{-a_k}\N _+^{a_k}) \N _{-a_{k-1}}\ldots  \N _{-a_1} \N _+^{a_1}\ldots  \N _+^{a_{k-1}}\nonumber{} \\  
 &\quad  - (k-1) (4\p _+) \N _{-a_{k-1}}\ldots  \N _{-a_1} \N _+^{a_1}\ldots  \N _+^{a_{k-1}}\label{relation-1} \end{align}
This equation is used to prove the second one
\begin{align}  
 &\N _{-a_n}\ldots  \N _{-a_1}\N _+^{a_1}\ldots  \N _+^{a_n} (\N _+\s ^{kl}\N _+) \;=\;\nonumber{} \\ \;=\;
 &(\N _+\s ^{kl}\N _+) \left ( \sum _{i=0}^n C^{(n)}_i (4\p _+)^{n-i} \N _{-a_i}\ldots  \N _{-a_1}\N _+^{a_1}\ldots \N _+^{a_i} \right )\label{relation-2} \end{align}
where the numeric coefficients $C^{(n)}_i$ with $i=0,\ldots ,n$ are found recursively starting with
  \begin{equation}
   C^{(1)}_0 = -2\  ,\quad  C^{(1)}_1 = 1
   \end{equation}
and the following ones with
\begin{equation}
   \begin{aligned}     
   C^{(n+1)}_0 & = -(n+2) C^{(n)}_0 \\
   C^{(n+1)}_i & = (i-n-2)C^{(n)}_i + C^{(n)}_{i-1}\  ,\quad  i=1,\ldots ,n \\
   C^{(n+1)}_{n+1} & = C^{(n)}_n
   \end{aligned}
   \end{equation}
Using equations (\ref{relation-1}) and (\ref{relation-2}) continuously we can finally arrive at the expressions needed to compute the picture-zero vertex operator
  \begin{align}
  \N _{-a_2}\N _{-a_1}(\N _+\s ^{ij}\N _+)(\N _+\s ^{kl}\N _+)(\N _+\s ^{mn}\N _+)\N _+^{a_1}\N _+^{a_2} \Phi & = \nonumber{} \\
  \;=\; 2! (4\p _+)^2 (\N _+\s ^{ij}\N _+)(\N _+\s ^{kl}\N _+)&(\N _+\s ^{mn}\N _+)\Phi\label{expr-1} \\
  \N _{-a_4}\ldots \N _{-a_1}(\N _+\s ^{ij}\N _+)(\N _+\s ^{kl}\N _+)\N _+^{a_1}\ldots \N _+^{a_4}\Phi & = \nonumber{} \\
  \;=\;  4! (4\p _+)^4 (\N _+\s ^{ij}\N _+)&(\N _+\s ^{kl}\N _+)\Phi\label{expr-2} \\
  \N _{-a_6}\ldots  \N _{-a_1}(\N _+\s ^{ij}\N _+)\N _+^{a_1}\ldots \N _+^{a_6}\Phi & = \nonumber{} \\
  \;=\; 6!(4\p _+)^6 & (\N _+\s ^{ij}\N _+)\Phi\label{expr-3}
  \end{align}

\acknowledgments

We want to thank Nathan Berkovits, Lucas Martins and Michael Movshev for discussions.
This work was supported
in part by FAPESP grant 2014/18634-9 ``Dualidade Gravitac$\!\!,\tilde{\rm a}$o/Teoria de Gauge''.


\begin{thebibliography}{10}

\bibitem{Berkovits:2008ga}
N.~Berkovits, {\it {Simplifying and Extending the $AdS_5 \times S^5$ Pure Spinor
  Formalism}},  {\em JHEP} {\bf 09} (2009) 051 doi: {\bf
  10.1088/1126-6708/2009/09/051}, \href{https://arxiv.org/abs/0812.5074}{arXiv:0812.5074 [hep-th]}.

\bibitem{Berkovits:2008qc}
N.~Berkovits, {\it {Perturbative Super-Yang-Mills from the Topological
  $AdS_5 \times S^5$ Sigma Model}},  {\em JHEP} {\bf 09} (2008) 088 doi: {\bf
  10.1088/1126-6708/2008/09/088}, \href{https://arxiv.org/abs/0806.1960}{arXiv:0806.1960 [hep-th]}.

\bibitem{Berkovits:2000yr}
N.~Berkovits and O.~Chandia, {\it Superstring vertex operators in an $AdS_5 \times S^5$ background},  {\em Nucl. Phys.} {\bf B596} (2001) 185--196  doi: {\bf 10.1016/S0550-3213(00)00697-0}, \href{https://arxiv.org/abs/hep-th/0009168}{arXiv:hep-th/0009168}.

\bibitem{Mikhailov:2011si}
A.~Mikhailov, {\it {Symmetries of massless vertex operators in $AdS_5 \times S^5$}},
   {\em Adv.Theor.Math.Phys.} {\bf 15} (2011) 1319--1372 doi: {\bf
  10.4310/ATMP.2011.v15.n5.a3}, \href{https://arxiv.org/abs/0903.5022}{arXiv:0903.5022 [hep-th]}.

\bibitem{Bedoya:2010qz}
O.~A. Bedoya, L.~Bevilaqua, A.~Mikhailov, and V.~O. Rivelles, {\it {Notes on
  beta-deformations of the pure spinor superstring in $AdS_5 \times S^5$}},  {\em
  Nucl.Phys.} {\bf B848} (2011) 155--215 doi: {\bf
  10.1016/j.nuclphysb.2011.02.012}, \href{https://arxiv.org/abs/1005.0049}{arXiv:1005.0049 [hep-th]}.

\bibitem{Berkovits:2019rwq}
N.~Berkovits, {\it {Half-BPS vertex operators of the AdS$_{5} \times$ S$^{5}$
  superstring}},  {\em JHEP} {\bf 07} (2019) 084 doi: {\bf
  10.1007/JHEP07(2019)084},  \href{https://arxiv.org/abs/1904.06564}{arXiv:1904.06564 [hep-th]}.

\bibitem{Grassi:2004tv}
P.~A. Grassi and G.~Policastro, {\it {Super-Chern-Simons theory as superstring
  theory}}, \href{https://arxiv.org/abs/hep-th/0412272}{arXiv:hep-th/0412272}.

\bibitem{Berkovits:2006vi}
N.~Berkovits and N.~Nekrasov, {\it {Multiloop superstring amplitudes from
  non-minimal pure spinor formalism}},  {\em JHEP} {\bf 0612} (2006) 029 doi:
  {\bf 10.1088/1126-6708/2006/12/029}, \href{https://arxiv.org/abs/hep-th/0609012}{arXiv:hep-th/0609012}.

\bibitem{Cremonini:2019aao}
   C.A.~Cremonini and P.A.~Grassi, {\it {Pictures from Super Chern-Simons Theory}},
   {\em JHEP} {\bf 03} (2020) 043 doi:
   {\bf 10.1007/JHEP03(2020)043}, \href{https://arxiv.org/abs/1907.07152}{arXiv:1907.07152 [hep-th]}.

\bibitem{Martins:2019wwi}
L.~N.~S. Martins, {\it {Type IIB superstring vertex operator from the -8
  picture}}, \href{https://arxiv.org/abs/1912.06498}{1912.06498 [hep-th]}.

\bibitem{Friedan:1985ge}
D.~Friedan, E.~J. Martinec, and S.~H. Shenker, {\it {Conformal Invariance,
  Supersymmetry and String Theory}},  {\em Nucl. Phys.} {\bf B271} (1986)
  93--165 doi: {\bf 10.1016/0550-3213(86)90356-1,
  10.1016/S0550-3213(86)80006-2}.

\bibitem{Belopolsky:1997bg}
A.~Belopolsky, {\it {New geometrical approach to superstrings}}, \href{https://arxiv.org/abs/hep-th/9703183}{arXiv:hep-th/9703183}.

\bibitem{Belopolsky:1997jz}
A.~Belopolsky, {\it {Picture changing operators in supergeometry and
  superstring theory}}, \href{https://arxiv.org/abs/hep-th/9706033}{arXiv:hep-th/9706033}.

\bibitem{Movshev:2012gc}
M.~V. Movshev, {\it {The odd twistor transform in eleven-dimensional
  supergravity}}, \href{https://arxiv.org/abs/1206.0057}{arXiv:1206.0057 [hep-th]}.


\bibitem{Green:1983hw}
M.~B. Green, J.~H. Schwarz, and L.~Brink, {\it {Superfield Theory of Type II
  Superstrings}},  {\em Nucl. Phys. B} {\bf 219} (1983) 437--478 doi: {\bf
  10.1016/0550-3213(83)90651-X}.


\bibitem{Deligne:1999qp}
P.~Deligne, P.~Etingof, D.~S. Freed, L.~C. Jeffrey, D.~Kazhdan, J.~W. Morgan,
  D.~R. Morrison, and E.~Witten, eds., {\em {Quantum fields and strings: A
  course for mathematicians. Vol. 1, 2}},
\newblock AMS, Providence USA (1999).

\end{thebibliography}

\end{document}